\documentclass[fleqn,twoside]{article}

\usepackage{amssymb}
\usepackage{graphicx}

\usepackage{amsmath}
\numberwithin{equation}{section}

\begin{document}
\noindent
{\sf University of Shizuoka}

\hspace*{13cm} {\large US-04-02}
\vspace{3mm}
%\draft

\begin{center}
%\title{
{\Large\bf Neutrino Masses Induced by $R$-Parity Violation \\[.1in]
in a SUSY SU(5) Model with  Additional
$\bar{5}'_L+5'_L$} 

\vspace{3mm}

%\author{
{\bf Yoshio Koide}

%\address{
{\it Department of Physics, University of Shizuoka, 
52-1 Yada, Shizuoka 422-8526, Japan\\
E-mail address: koide@u-shizuoka-ken.ac.jp}

\vspace{2mm}
\date{\today}
\end{center}

\vspace{3mm}
%\maketitle
\begin{abstract}
Within the framework of an SU(5) SUSY GUT model, a possible 
general form of the neutrino mass matrix induced by 
$R$-parity violation is investigated.  The model has matter 
fields $\overline{5}'_{L}+5'_{L}$ in addition to the ordinary 
matter fields $\overline{5}_{L}+10_{L}$ and Higgs fields 
$H_u+\overline{H}_d$.  The $R$-parity violating terms are 
given by $\overline{5}_{L} \overline{5}_{L} 10_{L}$, while 
the Yukawa interactions are given by $\overline{H}_d 
\overline{5}'_{L} 10_{L}$.  Since the matter fields 
$\overline{5}'_L$ and $\overline{5}_L$ are different from 
each other at the unification scale, the $R$-parity violation 
effects at a low energy scale appear only through the 
$\overline{5}'_L \leftrightarrow \overline{5}_{L}$ mixings.
In order to make this $R$-parity violation effect harmless
for proton decay, a discrete symmetry Z$_3$ and a triplet-doublet 
splitting mechanism analogous to that in the 5-plet Higgs 
fields are assumed.
\end{abstract}

%\pacs{

%\maketitle

%%%%%%%%%%%%%%%%%%%%%%%%%%%%%%%%%%%%%%%%%%%%%%%%%%%%%%%%%%%%%
%\begin{multicols}{2}

%\narrowtext

\section{Introduction}
\label{Sec1}

As an origin of the neutrino masses, 
the idea of the radiative neutrino mass \cite{Zee} is very 
interesting as well as the idea of the neutrino seesaw mechanism 
\cite{seesaw}.
However, currently, the latter idea is influential, because it is 
hard to embed the former model into a grand unification theory (GUT). 
For example, a supersymmetric (SUSY) model with $R$-parity violation
can provide radiative neutrino masses \cite{R_SUSY}, 
but the model cannot be 
embedded into GUT, because the $R$-parity violating terms induce 
proton decay inevitably \cite{Smirnov}.

Recently, the author \cite{SUSY-Koide} has proposed a model with 
$R$-parity violation within the framework of an SU(5) SUSY GUT:
we have quark and lepton fields $\overline{5}_L +10_L$, which
contribute to the Yukawa interactions as $H_u 10_L 10_L$ and
$\overline{H}_d \overline{5}_L 10_L$; we also have additional 
matter fields $\overline{5}'_L + 5'_L$ which contribute to 
the $R$-parity violating terms $\overline{5}'_L \overline{5}'_L 10_L$.
Since the two $\overline{5}_L$ and $\overline{5}'_L$ are different
from each other, the $R$-parity violating interactions are usually
invisible.
The $R$-parity violating effects become visible only through 
$\overline{5}_L \leftrightarrow \overline{5}'_L$ mixings
in low energy phenomena.

In the previous model \cite{SUSY-Koide}, a discrete symmetry 
Z$_3$ has been assumed,
and their quantum numbers have been assigned as 
$\overline{5}_{L(-)} +10_{L(+)} +\overline{5}'_{L(+)} +5'_{L(+)}$
and $\overline{H}_{d(0)} +H_{u(+)}$, where we have denoted
fields with the transformation properties
$\Psi \rightarrow \omega^{+1} \Psi$,
$\Psi \rightarrow \omega^{0} \Psi$ and
$\Psi \rightarrow \omega^{-1} \Psi$ ($\omega= e^{i2\pi/3}$) as
$\Psi_{(+)}$, $\Psi_{(0)}$ and $\Psi_{(-)}$, respectively.
Therefore, in the set $\overline{5}_L +10_{L}$,
the fields $\overline{5}_{L(-)}$ and $10_{L(+)}$ have
different transformation properties each other.
In contrast to the previous model,
in the present paper, we will propose a model with alternative
assignments
\begin{equation}
(\overline{5}_{L} +10_{L})_{(+)} + (\overline{5}'_{L} +5'_{L})_{(0)}
+\overline{H}_{d(-)} +H_{u(+)} \ .
%(1.1)
\end{equation}
Although the mechanism of the harmless $R$-parity violation is the
same as the previous model, since the Z$_3$ quantum number
assignment is different from the previous one, the structure of
the model is completely different from the previous one.

In the present paper, we will investigate not only the 
radiatively-induced neutrino masses, but also the contributions
from the vacuum expectation values (VEV) of the sneutrinos,
$\langle \widetilde{\nu} \rangle$, although in the previous
paper the estimate of $\langle \widetilde{\nu} \rangle$ 
was merely based on an optimistic speculation.

%%%%%%%%%%%%%%%%%%%%%%%%%%%%%%%%%%%%%%%%%%%%%%%%%%%%%%%%%%%%%%%%%
\section{Harmless $R$-parity violation mechanism}
\label{Sec2}

Under the Z$_3$ quantum number assignment (1.1),
the Z$_3$ invariant tri-linear terms in the superpotential
are only the following three terms:
\begin{equation}
W_{tri}  =  (Y_u)_{ij} H_{u(+)} 10_{L(+)i} 10_{L(+)j} 
+ (Y_d)_{ij} \overline{H}_{d(-)}
\overline{5}'_{L(0)i}10_{L(+)j} +\lambda_{ijk} \overline{5}_{L(+)i}
\overline{5}_{L(+)j} 10_{L(+)k} \ .
%\eqno(2.1)
\end{equation}
%Here and hereafter, for convenience, we drop the prime in
%$\overline{5}'_{L(0)}$ and $5'_{L(0)}$.
Similarly, the Z$_3$ invariant bi-linear terms are only two:
$\overline{H}_{d(-)} H_{u(+)}$ and $\overline{5}_{L(0)} H_{L(0)}$.
In order to give doublet-triplet splitting,
we assume the following ``effective" bi-linear terms
\begin{equation}
W_{bi} =  \overline{H}_{d(-)}(\mu +g_H \langle\Phi_{(0)}\rangle) H_{u(+)}
%\nonumber \\
+ \overline{5}'_{L(0)i}(M_5 -g_5 \langle\Phi_{(0)}\rangle) 5'_{L(0)i}
%\nonumber \\
+  M_i^{SB} \overline{5}_{L(+)i}  5'_{L(0)i}  \ ,
%\eqno(2.2)
\end{equation}
where $\Phi_{(0)}$ is a 24-plet Higgs field with the VEV
$\langle \Phi_{(0)} \rangle= v_{24} {\rm diag}(2,2,2,-3,-3)$,
so that, for example, the effective masses $M^{(a)}$ in the term
$\overline{5}_{L(0)}^{\prime (a)} 5_{L(0)}^{\prime (a)}$  
($5^{(2)}_L$ and
$5^{(3)}_L$ denote doublet and triplet components of the
fields $5_L$, respectively) are given by
\begin{equation}
M^{(2)} = M_5 +3 g_5 v_{24} \ , \ \ \ 
M^{(3)} = M_5 -2 g_5 v_{24} \ .
%\eqno(2.3)
\end{equation}
The last term in Eq.~(2.2) has been added in order to break the
Z$_3$ symmetry softly.
We define the $\overline{5}_L \leftrightarrow \overline{5}'_L$
mixing as follows:
\begin{eqnarray}
\overline{5}'_{L(0)i} &=& c_i \overline{5}^{q\ell}_{Li} +  
s_i \overline{5}^{heavy}_{Li}  \ ,
\nonumber \\
\overline{5}_{L(+)i} &=& -s_i \overline{5}^{q\ell}_{Li}  +  
c_i \overline{5}^{heavy}_{Li}  \ ,
%\eqno(2.4)
\end{eqnarray}
where $s_i=\sin\theta_i$ and $c_i=\cos\theta_i$. 
Then, we can rewrite the second and third terms in Eq.~(2.2) as
\begin{equation}
 \sum_{a=2,3} \sqrt{ (M^{(a)})^2 +(M_i^{SB})^2}\,
\left(\overline{5}^{heavy}_{Li}\right)^{(a)} 
\left(5^{heavy}_{Li}\right)^{(a)}  \ ,
%\eqno(2.5)
\end{equation}
where $5^{heavy}_L = 5'_{L(0)}$ and
\begin{equation}
s_i^{(a)} = \frac{ M^{(a)} }{ \sqrt{ (M^{(a)})^2 +
(M_i^{SB})^2} } \ ,
\ \ \ 
c_i^{(a)} = \frac{ M_i^{SB} }{ \sqrt{ (M^{(a)})^2 
+(M_i^{SB})^2} } \ .
%\eqno(2.6)
\end{equation}
The fields $\overline{5}_{Li}^{heavy(a)}$ have  masses
$\sqrt{ (M^{(a)})^2 +(M_i^{SB})^2}$, while 
$\overline{5}_{Li}^{q\ell(a)}$ are massless.
We regard $\overline{5}^{q\ell}_{Li} + 10_{L(+)i}$ as the observed 
quarks and leptons at low energy scale ($\mu <M_{GUT}$).
(Hereafter, we will simply denote $\overline{5}_{Li}^{q\ell}$
and $10_{L(+)i}$ as $\overline{5}_{Li}$ and $10_{Li}$,
respectively.)

Then, the effective $R$-parity violating terms at $\mu< M_{GUT}$ 
are given by
\begin{equation}
W_{\not\!R}^{eff} = s_i^{(a)} s_j^{(b)} \lambda_{ijk} 
\overline{5}^{(a)}_{Li}\overline{5}^{(b)}_{Lj} 10_{Lk} \ .
%\eqno(2.7)
\end{equation}
In order to suppress the unwelcome term $d_R^c d_R^c u_R^c$ 
in the effective $R$-parity violating terms (2.7),
we assume a fine tuning
\begin{equation}
M^{(2)} \sim M_{GUT}, \ \ M^{(3)} \sim m_{SUSY}, \ \ 
M_i^{SB} \sim M_{GUT} \times 10^{-1},
%\eqno(2.8)
\end{equation}
where $m_{SUSY}$ denotes a SUSY breaking scale 
($m_{SUSY} \sim 1$ TeV),  so that
\begin{equation}
s_i^{(2)} = 1 -O(10^{-2})\ , \ \ 
c_i^{(2)} \simeq \frac{ M_i^{SB} }{ M^{(2)} }\sim  10^{-1}
\ ; \ \ \ \ 
s_i^{(3)} \simeq \frac{ M^{(3)} }{M_i^{SB} } \sim 10^{-12} 
\ , \ \ 
c_i^{(3)} = 1 -O(10^{-24})   \ . 
%\eqno(2.9)
\end{equation}
Note that in the present model the observed down-quarks 
$d_{Ri}^c=(\overline{5}_{Li}^{q\ell})^{(3)}$ are given by
$(\overline{5}_{Li}^{q\ell})^{(3)}\simeq (\overline{5}'_{L(0)i})^{(3)}$,
while the observed lepton doublets 
$(\nu_{Li},e_{Li})=(\overline{5}_{Li}^{q\ell})^{(2)}$ are given by
$(\overline{5}_{Li}^{q\ell})^{(2)}\simeq -(\overline{5}_{L(+)i})^{(2)}$.

{}From Eq.~(2.9), the $R$-parity violating terms  $d_R^c d_R^c u_R^c$ and 
$d_R^c (e_L u_L -\nu_L d_L)$ are suppressed by
$s^{(3)}s^{(3)}\sim 10^{-24}$ and $s^{(3)}s^{(2)}\sim 10^{-12}$,
respectively.
Thus, proton decay caused by terms  $d_R^c d_R^c u_R^c$ 
and $d_R^c (e_L u_L -\nu_L d_L)$ is suppressed by a factor
$(s^{(3)})^3s^{(2)}\sim 10^{-36}$.
On the other hand, radiative neutrino masses are generated
by the $R$-parity violating term $(e_L \nu_L -\nu_L e_L)e_R^c$
with a factor $s^{(2)}s^{(2)}\simeq 1$.
%The numerical choice (2.9) gives
%\begin{eqnarray}
%m(\overline{5}_{Li}^{\prime (2)})& \simeq & M^{(2)} \sim M_{GUT} \ ,
%\nonumber \\
%m(\overline{5}_{Li}^{\prime (3)})& \simeq & M^{SB}_i \sim 
%M_{GUT} \times 10^{-1} \ . 
%\eqno(1.11)
%\end{eqnarray}
%Since $m(\overline{5}_{Li}^{\prime (3)})< M_X$, the triplet
%fields $\overline{5}_{Li}^{\prime (3)}$ can basically contribute 
%to the renormalization group equation (RGE) effects at $\mu<M_X$.
%However, since we consider $M_i^{SB} \sim M_X\times 10^{-1}$, 
%the numerical effect does almost not spoil the gauge-coupling-constant
%unification at $\mu=M_X \sim 10^{16}$ GeV.

The up-quark masses are generated by the Yukawa interactions (2.1),
so that we obtain the up-quark mass matrix $M_u$ as 
%\begin{equation}
$(M_u)_{ij} = (Y_u)_{ij} v_u$,
%\eqno(1.12)
%\end{equation}
where $v_u=\langle H_{u(+)}^0\rangle$.
We also obtain the
down-quark mass matrix $M_d$ and charged lepton mass
matrix $M_e$ as
\begin{equation}
M_d^\dagger= C^{(3)} Y_d v_d\, \ \ \ 
M_e^* = C^{(2)} Y_d v_d \ , 
%\eqno(2.10)
\end{equation}
where
\begin{equation}
C^{(a)} = {\rm diag}(c_1^{(a)}, c_2^{(a)},c_3^{(a)}) \ ,
%(2.11)
\end{equation}
so that
\begin{equation}
M_d^T= \left(C^{(3)} C^{(2) -1}\right)^* M_e \ ,
%\eqno(2.12)
\end{equation}
where $v_d=\langle \overline{H}_{d(-)}^0\rangle$.
Note that $M_d^T$ has a structure different from $M_e$, 
because the values of $c_i^{(2)}$ ($i=1,2,3$)
can be different from each other.
(The idea  $M_d^T \neq M_e$ based on a mixing between two
$\overline{5}_L$ has been discussed, for example, by
Bando and Kugo \cite{two5mix} in the context of an E$_6$ model.)

%%%%%%%%%%%%%%%%%%%%%%%%%%%%%%%%%%%%%%%%%%%%%%%%%%%%%%%
%\vspace{5mm}
\section{General form of the neutrino mass matrix} 
\label{sec:3}

First, we investigate a possible form of the 
radiatively-induced neutrino mass matrix $M_{rad}$.
In the present model, since we do not have a term which induces
$\widehat{e}^+_R \leftrightarrow \overline{H}^+_{d(-)}$ mixing, 
there is no Zee-type diagram \cite{Zee}, 
which is proportional to the Yukawa vertex 
$(Y_d)_{ij}$ and R-parity violating vertex $\lambda_{ijk}$ .
%(The $\widehat{e}^+_R \leftrightarrow \overline{H}^+_{d(-)}$ mixing
%can come from interactions of a type $\overline{H}_d\, \overline{H}_d\,
%10_{L}$.  However, the combination $\overline{H}_{d(-)}
%\overline{H}_{d(-)}10_{L(+)}$ is forbidden because of 
%the antisymmetric property
%of SU(5) 10-plet fields $10_{L(+)}$.
%Even after the SU(5) is broken, 
%$\overline{H}_{d}^{(2)}\overline{H}_{d}^{(2)}$ cannot
%couple to the SU(2) singlet $\widehat{e}^+_R$ because
%SU(2) singlet composed of $2\times 2$ must be antisymmetric.
%Therefore, we cannot bring the $\overline{H}_{d}^{(2)}
%\overline{H}_{d}^{(2)}\widehat{e}^+_R$ term even as a
%soft supersymmetry breaking term.)

%%%%%%%%%%%%%%%%%%%%%%%%%%%%%%%%%%%%%%%%%%%%%%%%%%%%%%%%%%%%%%%%%%%%%%
%%%%% penguin %%%%%%%%%%%%%%%%%%%%%%%%%%%%%%%%%%%%%%%%%%%%%%%%%%%%%%%%
%%%%%%%%%%%%%%%%%%%%%%%%%%%%%%%%%%%%%%%%%%%%%%%%%%%%%%%%%%%%%%%%%%%%%%
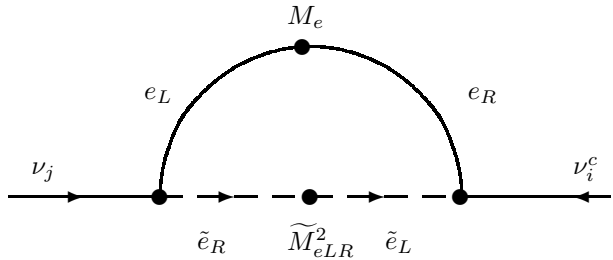
\begin{figure}[hb]
\unitlength=1cm
%\begin{picture}(16.6,4)
\begin{center}
\begin{picture}(9,4)
\thicklines
%
% lepton type
%
% nu_i
\put(0.5,1){\line(1,0){2}}
\put(1.5,1){\vector(1,0){0}}
\put(0.8,1.3){$\nu_j$}
% vertex
\put(2.5,1){\circle*{0.2}}
% slepton
\multiput(2.5,1)(0.5,0){8}{\line(1,0){0.3}}
\put(3.5,1){\vector(1,0){0}}
\put(3,0.3){$\tilde{e}_R$}
% vertex
\put(4.5,1){\circle*{0.2}}
% slepton
\put(5.5,1){\vector(1,0){0}}
\put(5.5,0.3){$\tilde{e}_L$}
% vertex
\put(6.5,1){\circle*{0.2}}
\put(4.2,0.3){$\widetilde{M}^{2}_{eLR}$}
% neutrino
\put(6.5,1){\line(1,0){2}}
\put(8,1){\vector(-1,0){0}}
\put(8,1.3){$\nu_i^c$}
%%%% charged lepton circle %%%%%
\put(4.5,1){
\qbezier(-2,0)(-2.01,0.35)(-1.88,0.68)
\qbezier(-1.88,0.68)(-1.78,1.03)(-1.53,1.29)
\qbezier(-1.53,1.29)(-1.3,1.55)(-1,1.73)
\qbezier(-1,1.73)(-0.69,1.9)(-0.35,1.97)
\qbezier(-0.35,1.97)(0,2.03)(0.35,1.97)
\qbezier(2,0)(2.01,0.35)(1.88,0.68)
\qbezier(1.88,0.68)(1.78,1.03)(1.53,1.29)
\qbezier(1.53,1.29)(1.3,1.55)(1,1.73)
\qbezier(1,1.73)(0.69,1.9)(0.35,1.97)
}
\put(4.4,3){\circle*{0.2}}
\put(4.2,3.3){$M_e$}
\put(2.3,2.3){$e_L$}
\put(6.6,2.3){$e_R$}
%
% down type
%
% nu_i
%\put(8.8,1){\line(1,0){2}}
%\put(9.8,1){\vector(1,0){0}}
%\put(9.1,1.3){$\nu_j$}
% vertex
%\put(10.8,1){\circle*{0.2}}
% slepton
%\multiput(10.8,1)(0.5,0){8}{\line(1,0){0.3}}
%\put(11.8,1){\vector(1,0){0}}
%\put(11.3,0.3){$\tilde{d}_R$}
% vertex
%\put(12.8,1){\circle*{0.2}}
% slepton
%\put(13.8,1){\vector(1,0){0}}
%\put(13.8,0.3){$\tilde{d}_L$}
% vertex
%\put(14.8,1){\circle*{0.2}}
%\put(12.5,0.3){$\widetilde{M}^{2}_{dLR}$}
% neutrino
%\put(14.8,1){\line(1,0){2}}
%\put(16.3,1){\vector(-1,0){0}}
%\put(16.3,1.3){$\nu_i^c$}
%%%% downquark circle %%%%%
%\put(12.8,1){
%\qbezier(-2,0)(-2.01,0.35)(-1.88,0.68)
%\qbezier(-1.88,0.68)(-1.78,1.03)(-1.53,1.29)
%\qbezier(-1.53,1.29)(-1.3,1.55)(-1,1.73)
%\qbezier(-1,1.73)(-0.69,1.9)(-0.35,1.97)
%
%\qbezier(-0.35,1.97)(0,2.03)(0.35,1.97)
%
%\qbezier(2,0)(2.01,0.35)(1.88,0.68)
%\qbezier(1.88,0.68)(1.78,1.03)(1.53,1.29)
%\qbezier(1.53,1.29)(1.3,1.55)(1,1.73)
%\qbezier(1,1.73)(0.69,1.9)(0.35,1.97)
%}
%\put(12.7,3){\circle*{0.2}}
%\put(12.5,3.3){$M_d$}
%\put(10.6,2.3){$d_L$}
%\put(14.9,2.3){$d_R$}
\end{picture}
\caption{Radiative generation of neutrino Majorana mass}
\label{fig:numass}
\end{center}
\end{figure}
%%%%%

Only the radiative neutrino masses in the present scenario come
from a charged-lepton loop diagram:
the radiative diagram with 
$(\nu_L)_j \rightarrow (e_R)_l + (\widetilde{e}_L^c)_n$ and
$(e_L)_k + (\widetilde{e}_L^c)_m \rightarrow (\nu_L^c)_i$.
The contributions $(M_{rad})_{ij}$ from the charged lepton 
loop are given,
except for the common factors, as follows:
%%%%%%%%%%%%%%%%%%%%%%%%%%%%%%%%%%%%%%%%%%%%%%%%%%%%%%%%%%%%%%%%%%
\begin{equation}
(M_{rad})_{ij}= 
s_i s_j s_k s_n \lambda^*_{ikm} \lambda^*_{jnl}
  (M_e)^*_{kl} (\widetilde{M}_{eLR}^{2T})^*_{mn} + 
(i \leftrightarrow j)  \ ,
%\eqno(3.1)
\end{equation}
where $s_i=s_i^{(2)}$, and 
$M_e$ and $\widetilde{M}_{eLR}^2$ are charged-lepton and 
charged-slepton-LR mass matrices, respectively.
(In the present paper, we define the charged lepton mass
matrix $M_e$ and the neutrino mass matrix $M_\nu$ as
$\overline{e}_L M_e e_R$ and $\overline{\nu}_L M_\nu \nu_L^c$,
respectively, so that the complex conjugate quantities 
$\lambda^*_{ijk}$ and so on have appeared in the expression (3.1).)
Since  $\widetilde{M}_{eLR}^2$ is proportional to $M_e$, i.e. 
$\widetilde{M}^2_{eLR} = (A+\mu^{(2)}\tan\beta)  M_e$
($\mu^{(2)} = \mu -3 g_H v_{24}$, and $A$ is the coefficient 
of the soft SUSY breaking terms
$(Y_d)_{ij} (\tilde{\nu}, \tilde{e})^T_{Li} \tilde{e}_{Lj}^c
\overline{H}_{d}$ with $A \sim 1$ TeV), we obtain
\begin{equation}
(M_{rad})_{ij} = 2(A+\mu^{(2)}\tan\beta)
s_i s_j s_k s_n \lambda^*_{ikm} \lambda^*_{jnl} (M_e)^*_{kl} (M_e)^*_{nm}
 \ .
%\eqno(3.2)
\end{equation}
Since the coefficient $\lambda_{ijk}$ is antisymmetric in the permutation
$i \leftrightarrow j$, it is useful to define
\begin{equation}
\lambda_{ijk} = \varepsilon_{ijl} L_{lk} \ , 
%\eqno(3.3)
\end{equation}
and
\begin{equation}
K =  (S M_e L^T)^* \ ,
%\eqno(3.4)
\end{equation}
where $S={\rm diag}(s_1,s_2,s_3)$.
Then, the radiative neutrino mass matrix is given by 
\begin{equation}
(M_{rad})_{ij} = m_0^{-1} s_i s_j \varepsilon_{ikm} 
\varepsilon_{jln} K_{ml} K_{nk} \ .
%\eqno(3.5)
\end{equation}
The coefficient $m_0^{-1}$ is calculated from one-loop diagram
(Fig.1) as
\begin{equation}
m_0^{-1} = \frac{2}{16 \pi^2} (A+\mu^{(2)}\tan\beta)
 F(m_{\tilde{e}_R}^2, m_{\tilde{e}_L}^2)  \ ,
%\eqno(3.6)
\end{equation}
where
\begin{equation}
F(m_a^2, m_b^2)= \frac{1}{m_a^2-m_b^2} \ln\frac{m_a^2}{m_b^2} \ .
%\eqno(3.7)
\end{equation}

%%%%%%%%%%%%%%%%%%%%%%%%%%%%%%%%%%%%%%%%%%%%%%%%%%%%%%%
%\vspace{5mm}
%\section{Contributions from the sneutrino VEV} 
%\label{sec:4}

Next, let us investigate the contributions from the VEVs of
sneutrinos $\langle \widetilde{\nu}_i \rangle $.
In general, the sneutrinos $\widetilde{\nu}_i$ 
can have VEVs 
$v_i \equiv \langle \widetilde{\nu}_i \rangle \neq 0$ \cite{snu},
if there are  one or more of the following terms: 
$\mu_i \overline{5}_{Li} H_u$ in superpotential $W$, 
and $B_i \overline{5}_{Li} H_u + m_{HLi}^2 \overline{5}_{Li} 
\bar{H}_d^\dagger$ 
in the bilinear soft SUSY breaking terms $V_{soft}$.
In the present model, there is no such a term at tree level,
because these terms are forbidden by the Z$_3$ symmetry.
However, only an effective $m_{HLi}^2$-term can appear 
via the loop diagram
$\overline{H}_d \rightarrow (\overline{5}^{ql}_L)^c + (10_L)^c 
\rightarrow \overline{5}_L^{ql}$ (Fig.~2).
The contribution $m_{HLi}^2$ is proportional to
\begin{equation}
s_i s_j \lambda_{ijk} (M_e)_{jk} =s_i \varepsilon_{ijk} K^*_{jk}
\ .
%(3.8)
\end{equation}
On the other hand, the contribution $M_{VEV}$ from $
\langle \widetilde{\nu}_i \rangle \neq 0$
to the neutrino mass matrix is proportional to
\begin{equation}
\left(
\begin{array}{ccc}
v^2_1 & v_1 v_2 & v_1 v_3 \\
v_1 v_2 & v^2_2 & v_2 v_3 \\
v_1 v_3 & v_2 v_3 & v_3^2 
\end{array} \right) 
\ ,
%\eqno(3.9)
\end{equation}
and $v_i \equiv \langle \widetilde{\nu}_i \rangle$
are proportional to the values $(m_{HLi}^2)^*$,
so that 
the mass matrix $M_{VEV}$ is given by
\begin{equation}
(M_{VEV})_{ij} =  \xi m_0^{-1} s_i s_j \varepsilon_{ikl} 
\varepsilon_{jmn} K_{kl} K_{mn} \ ,
%\eqno(3.10)
\end{equation}
where $\xi$ is a relative ratio of $M_{VEV}$ to
$M_{rad}$.

%%%%%%%%%%%%%%%%%%%%%%%%%%%%%%%%%%%%%%%%%%%%%%%%%%%%%%%%%%%%%%%%%%%%%%
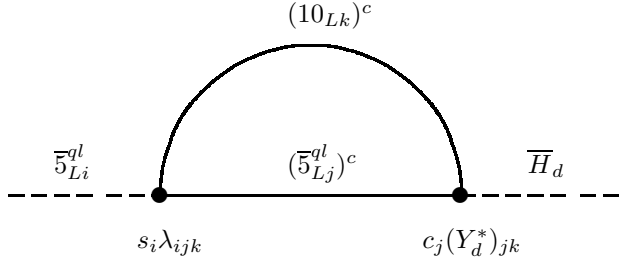
\begin{figure}[hb]
\unitlength=1cm
%\begin{picture}(16.6,4)
\begin{center}
\begin{picture}(9,4)
\thicklines
%
% boson_left
%
\put(0.5,1){\line(1,0){0.2}}
\put(0.8,1){\line(1,0){0.2}}
\put(1.1,1){\line(1,0){0.2}}
\put(1.4,1){\line(1,0){0.2}}
%\put(1.5,1){\vector(1,0){0}}
\put(1.1,1){\line(1,0){0.2}}
\put(1.8,1){\line(1,0){0.2}}
\put(1.1,1.3){$\overline{5}^{ql}_{Li}$}
\put(2.1,1){\line(1,0){0.2}}
\put(2.4,1){\line(1,0){0.2}}
% vertex
\put(2.5,1){\circle*{0.2}}
% line 5bar
%\multiput(2.5,1)(0.5,0){8}{\line(1,0){0.3}}
\put(2.5,1){\line(1,0){4}}
\put(4.2,1.3){$(\overline{5}^{ql}_{Lj})^c$}
% vertex
%\put(4.5,1){\circle*{0.2}}
% slepton
%\put(5.5,1){\vector(1,0){0}}
%\put(5.5,0.3){$\tilde{e}_L$}
% vertex
\put(6.5,1){\circle*{0.2}}
%\put(4.2,0.3){$\widetilde{M}^{2}_{eLR}$}
% boson_right
\put(6.5,1){\line(1,0){0.2}}
\put(6.8,1){\line(1,0){0.2}}
\put(7.1,1){\line(1,0){0.2}}
\put(7.4,1){\line(1,0){0.2}}
%\put(1.5,1){\vector(1,0){0}}
\put(7.4,1.3){$\overline{H}_d$}
\put(7.7,1){\line(1,0){0.2}}
\put(8.1,1){\line(1,0){0.2}}
\put(8.4,1){\line(1,0){0.2}}

%\put(6.5,1){\line(1,0){2}}
%\put(8,1){\vector(-1,0){0}}
%\put(8,1.3){$\overline{H}_d$}
%%%% 10_L  circle %%%%%
\put(4.5,1){
\qbezier(-2,0)(-2.01,0.35)(-1.88,0.68)
\qbezier(-1.88,0.68)(-1.78,1.03)(-1.53,1.29)
\qbezier(-1.53,1.29)(-1.3,1.55)(-1,1.73)
\qbezier(-1,1.73)(-0.69,1.9)(-0.35,1.97)
\qbezier(-0.35,1.97)(0,2.03)(0.35,1.97)
\qbezier(2,0)(2.01,0.35)(1.88,0.68)
\qbezier(1.88,0.68)(1.78,1.03)(1.53,1.29)
\qbezier(1.53,1.29)(1.3,1.55)(1,1.73)
\qbezier(1,1.73)(0.69,1.9)(0.35,1.97)
}
%\put(4.4,3){\circle*{0.2}}
\put(4.2,3.3){$(10_{Lk})^c$}
\put(2.2,0.3){$s_i \lambda_{ijk}$}
\put(6.0,0.3){$c_j(Y_d^*)_{jk}$}
%
%%%%%%%%%%%%%%%
%\put(4.5,1){
%\qbezier(-2,0)(-2.01,-0.35)(-1.88,-0.68)
%\qbezier(-1.88,-0.68)(-1.78,-1.03)(-1.53,-1.29)
%\qbezier(-1.53,-1.29)(-1.3,-1.55)(-1,-1.73)
%\qbezier(-1,-1.73)(-0.69,-1.9)(-0.35,-1.97)
%
%\qbezier(-0.35,-1.97)(0,-2.03)(0.35,-1.97)
%
%\qbezier(2,0)(2.01,-0.35)(1.88,-0.68)
%\qbezier(1.88,-0.68)(1.78,-1.03)(1.53,-1.29)
%\qbezier(1.53,-1.29)(1.3,-1.55)(1,-1.73)
%\qbezier(1,-1.73)(0.69,-1.9)(0.35,-1.97)
%}
%\put(4.4,-3){\circle*{0.2}}
%\put(4.2,-3.3){$M_e$}
%\put(2.3,-2.3){$e_L$}
%\put(6.6,-2.3){$e_R$}
% down type
%
% nu_i
%\put(8.8,1){\line(1,0){2}}
%\put(9.8,1){\vector(1,0){0}}
%\put(9.1,1.3){$\nu_j$}
% vertex
%\put(10.8,1){\circle*{0.2}}
% slepton
%\multiput(10.8,1)(0.5,0){8}{\line(1,0){0.3}}
%\put(11.8,1){\vector(1,0){0}}
%\put(11.3,0.3){$\tilde{d}_R$}
% vertex
%\put(12.8,1){\circle*{0.2}}
% slepton
%\put(13.8,1){\vector(1,0){0}}
%\put(13.8,0.3){$\tilde{d}_L$}
% vertex
%\put(14.8,1){\circle*{0.2}}
%\put(12.5,0.3){$\widetilde{M}^{2}_{dLR}$}
% neutrino
%\put(14.8,1){\line(1,0){2}}
%\put(16.3,1){\vector(-1,0){0}}
%\put(16.3,1.3){$\nu_i^c$}
%%%% downquark circle %%%%%
%\put(12.8,1){
%\qbezier(-2,0)(-2.01,0.35)(-1.88,0.68)
%\qbezier(-1.88,0.68)(-1.78,1.03)(-1.53,1.29)
%\qbezier(-1.53,1.29)(-1.3,1.55)(-1,1.73)
%\qbezier(-1,1.73)(-0.69,1.9)(-0.35,1.97)
%
%\qbezier(-0.35,1.97)(0,2.03)(0.35,1.97)
%
%\qbezier(2,0)(2.01,0.35)(1.88,0.68)
%\qbezier(1.88,0.68)(1.78,1.03)(1.53,1.29)
%\qbezier(1.53,1.29)(1.3,1.55)(1,1.73)
%\qbezier(1,1.73)(0.69,1.9)(0.35,1.97)
%}
%\put(12.7,3){\circle*{0.2}}
%\put(12.5,3.3){$M_d$}
%\put(10.6,2.3){$d_L$}
%\put(14.9,2.3){$d_R$}
\end{picture}
\caption{Effective $\overline{5}^{ql}_L \overline{H}_d^\dagger$ term}
\label{fig:vev}
\end{center}
\end{figure}
%%%%%

In conclusion, the neutrino mass matrix $M_\nu$ in the present
model is given by the form
\begin{equation}
(M_\nu)_{ij} =   m_0^{-1} s_i s_j \varepsilon_{ikl} 
\varepsilon_{jmn} \left(K_{kn} K_{ml} + \xi K_{kl} K_{mn} \right) \ ,
%\eqno(3.11)
\end{equation}
i.e.
\begin{eqnarray}
M_\nu &=&   m_0^{-1} S \left\{ \left[ 
(K -K^T)(K- K^T) - {\bf 1} {\rm Tr}(KK-KK^T) \right] (1+\xi) 
\right. \nonumber \\
 & & \left. + \left[ (K+K^T) -{\bf 1} {\rm Tr}K \right] {\rm Tr}K
-(KK+K^TK^T)+ {\bf 1} {\rm Tr}(KK) \right\} S \ ,
%\eqno(3.12)
\end{eqnarray}
where {\bf 1} is a $3\times 3$ unit matrix.
%The expression (3.11) is explicitly given as follows:
%\begin{eqnarray}
%(M_\nu)_{11}& =& m_0^{-1} s_1^2 \left[ (K_{23}-K_{32})^2 (1+\xi) 
%+2 (K_{23} K_{32}-K_{22} K_{33}) \right] \ , \nonumber \\
%(M_\nu)_{22} &=&  m_0^{-1} s_2^2 \left[ (K_{13}-K_{31})^2 (1+\xi) 
%+2 (K_{13} K_{31}-K_{11} K_{33}) \right] \ , \nonumber  \\
%(M_\nu)_{33} &=&  m_0^{-1} s_3^2 \left[ (K_{12}-K_{21})^2 (1+\xi) 
%+2 (K_{12} K_{21}-K_{11} K_{22}) \right] \ , \nonumber  \\
%(M_\nu)_{12} &=&  m_0^{-1} s_1 s_2 \left[ (K_{23}-K_{32})(K_{31}-K_{13}) 
%(1+\xi) + (K_{12} + K_{21})K_{33} - (K_{23} K_{31} +K_{32} K_{13})\right]
%\nonumber  \ , \\
%(M_\nu)_{13} &=&  m_0^{-1} s_1 s_3 \left[ (K_{23}-K_{32})(K_{12}-K_{21}) 
%(1+\xi) 
%+ (K_{13} + K_{31})K_{22} - (K_{23} K_{12} +K_{32} K_{21})\right] \ ,
%\nonumber  \\
%(M_\nu)_{23} &=&  m_0^{-1} s_2 s_3 \left[ (K_{31}-K_{13})(K_{12}-K_{21}) 
%(1+\xi) 
%+ (K_{23} + K_{32})K_{11} - (K_{31} K_{12} +K_{13} K_{21})\right] \ . 
%\nonumber \\
%\eqno(3.13)
%\end{eqnarray}

%%%%%%%%%%%%%%%%%%%%%%%%%%%%%%%%%%%%%%%
%\vspace{5mm}
\section{General features of the neutrino mass matrix} 
\label{sec:4}

In the present model, if the charged lepton mass matrix $M_e$ and the 
structure of $\lambda_{ijk}$ (i.e. $L_{ij}$) are given, then we can obtain 
$K=(S M_e L^T)^*$, so that we can predict neutrino masses and mixings.
However, at present, we have many unknown parameters, 
so that in order to give 
explicit predictions of the neutrino masses and mixings, 
we must put a further 
assumption on the parameters $K_{ij}$. 
In the present section, we investigate general features 
of the neutrino mass matrix (3.11) [or (3.12)] 
without making any explicit assumptions about flavor symmetries.

So far, the expression of $M_{\nu}$, \ (3.12), has been given 
in the initial
flavor basis, \ where $\overline{5}_{L(+)} \leftrightarrow 
\overline{5}^{'}_{L(0)}$ mixings have been taken place a diagonal form
\begin{equation}
S^{(a)} = {\rm diag} (s^{(a)}_1 , s^{(a)}_2 , s^{(a)}_3 ) \  , \ \ \ 
C^{(a)} = {\rm diag} (c^{(a)}_1 , c^{(a)}_2 , c^{(a)}_3 ) \  ,
%\eqno(4.1)
\end{equation}
and the matrix $K$ has been defined by Eq.~(3.4), $K=(S M_e L^T)^{*}$. 
Since $S$,
$M_e$ and $L$ are transformed as 
\begin{eqnarray}
M_e &\rightarrow& M'_e = U_5^{\dagger} M_e U^{*}_{10} \  , \nonumber \\
L   &\rightarrow& L'   = U_5^{\dagger} L   U_{10} \  , \\
S   &\rightarrow& S'   = U_5^{\dagger} S   U_{5} \  , \nonumber
%\eqno(4.2)
\end{eqnarray}
under a rotation of the flavor basis
\begin{equation}
10_L \rightarrow 10'_L = U_{10}^{\dagger} 10_L \  , \ \ \ 
\overline{5}_L^{ql} \rightarrow (\overline{5}_L^{q\ell})' 
= U_5^{\dagger} \overline{5}_L^{q\ell} \  , 
%\eqno(4.3)
\end{equation}
the matrix $K$ transforms as 
\begin{equation}
K \rightarrow K' = U_{5}^{T} \ K \ U_5 \ .
%\eqno(4.4)
\end{equation}
We have a great interest in the form of $M'_\nu$ in the flavor basis
with $M'_e=D_e \equiv {\rm diag}(m_e, m_\mu, m_\tau)$.
Hereafter, we denote the quantities $M'_\nu$, $K'$, and so on
in the $M'_e=D_e$ basis as $\widehat{M}_\nu$, $\widehat{K}$ and so on, 
respectively.
The matrix $\widehat{K}$ is expressed as
\begin{equation}
\widehat{K}  = \widehat{S} D_e \widehat{L}^\dagger
\simeq D_e (U_{R}^e)^{\dagger} L^\dagger (U^e_L) \ ,
%\eqno(4.5)
\end{equation}
where $U_5 =U^e_L$ and $U_{10}=U^e_R$, and
we have put $\widehat{S} \simeq {\bf 1}$ because of 
$S \simeq {\bf 1}$ as we have assumed in Eq.~(2.9).

Here, let us summarize general features of the present neutrino
mass matrix (3.12).

\noindent{(i)}
If the matrix $K$ defined by Eq.~(3.4) satisfies $K^{T} = K$ in the initial
basis, the matrix $K^{\prime}$ in the arbitrary basis also satisfies 
$K^{\prime T} = K^{\prime}$, so that the present model gives 
$\langle \widetilde{\nu}_i^{\prime} \rangle =0$ in the arbitrary basis. 
For
such a case, the neutrino mass matrix is simply given by
\begin{equation}
M_{\nu} = - m^{-1}_0 S \left[ 2KK - 2K {\rm Tr} K - 
{\rm Tr} (KK) + ({\rm Tr} K)^2 \right] S
 \ .
%\eqno(4.6)
\end{equation}

\noindent{(ii)}
When $K$ is symmetric under the flavor $2 \leftrightarrow 3$ 
permutation, the 
neutrino mass matrix $M_{\nu}$ is also symmetric under the 
$2 \leftrightarrow 3$ permutation. 
It is well-known \cite{23sym} that 
when the neutrino mass matrix $\widehat{M}_{\nu}$ is symmetric 
under the $2 \leftrightarrow 3$
permutation, the mass matrix $\widehat{M}_{\nu}$ gives 
a nearly bimaximal mixing, 
i.e.
$\sin^2 2 \theta_{23} =1$ and $|U_{13}|^2 =0$,
which are favorable to the observed atmospheric \cite{atm},
K2K \cite{K2K} and CHOOZ \cite{CHOOZ} data. 
In the present model, the $2 \leftrightarrow 3$ symmetry of
$\widehat{M}_\nu$ means that the parameters 
\begin{equation}
\widehat{K}_{ij} = K_{kl} (U^e_L)_{ki} (U_L^e)_{lj}  \ ,
%\eqno(4.7)
\end{equation}
are symmetric under the $2 \leftrightarrow 3$ permutation. 
In other words, the
$2 \leftrightarrow 3$ symmetry of $\widehat{M}_\nu$ is 
due to special structures of $U^e_L$ and $K$. 
For example, when $K$ and $U^e_L$ are given by the textures
\begin{equation}
K=
\left(
\begin{array}{ccc}
K_{11} & 0 & 0 \\
0 & K_{22} & K_{23} \\
0 & K_{32} & K_{33} 
\end{array} \right) \ ,
%\eqno(4.8)
\end{equation}
\begin{equation}
U^e_L =
\left(
\begin{array}{ccc}
0 & -\frac{1}{\sqrt2} & \frac{1}{\sqrt2} \\
-s & \frac{1}{\sqrt2}c & \frac{1}{\sqrt2}c \\
c & \frac{1}{\sqrt2}s & \frac{1}{\sqrt2}s
\end{array} \right) \ ,
%\eqno(4.9)
\end{equation}
the matrix $\widehat{K}$ is $2 \leftrightarrow 3$ symmetric:
\begin{equation}
\widehat{K} =
\left(
\begin{array}{ccc}
f & a & a \\
a^{\prime} & g & b \\
a^{\prime} & b & g 
\end{array} \right) \ ,
%\eqno(4.10)
\end{equation}
so that the neutrino mass matrix $\widehat{M}_{\nu}$ is also 
$2 \leftrightarrow 3$ symmetric:
\begin{eqnarray}
(\widehat{M}_{\nu})_{11} &=& -2(g^2 - b^2) m_0^{-1} \  , \nonumber \\
(\widehat{M}_{\nu})_{12} &=& (\widehat{M}_{\nu})_{13} 
= (\widehat{M}_{\nu})_{21} = (\widehat{M}_{\nu})_{31} 
= (a + a^{\prime})(g - b) m_0^{-1} \  , \\
(M_{\nu})_{22} &=& (\widehat{M}_{\nu})_{33}  
= \left[(a - a^{\prime})^2 (1+\xi) +
2(a a^{\prime} - fg) \right] m_0^{-1} \  , \nonumber \\
(\widehat{M}_{\nu})_{23} &=& (\widehat{M}_{\nu})_{32} 
= - \left[(a - a^{\prime})^2 (1+\xi) + 2(a a^{\prime} - fb) 
\right] m_0^{-1}  \  , \nonumber
%\eqno(4.11)
\end{eqnarray}
and $K$ in the initial basis is given by 
\begin{eqnarray}
K_{11} &=& g - b \  , \nonumber \\
K_{22} &=& (g+b)c^2 - \sqrt{2} (a + a^{\prime})cs + fs^2 \  , \nonumber \\
K_{33} &=& (g+b)s^2 + \sqrt{2} (a + a^{\prime})cs + fc^2 \  , \\
K_{23} &=& \sqrt{2}(-as^2 + a^{\prime}c^2 ) + (g+b-f)cs \  , \nonumber \\
K_{32} &=& \sqrt{2}(ac^2 - a^{\prime}s^2 ) + (g+b-d)cs \ . \nonumber
%\eqno(4.12)
\end{eqnarray}

Finally, let us show a simple example which is suggested by 
above comments (i) and (ii). 
We assume that $M_e M_e^{\dagger}$ on the initial 
basis is $2 \leftrightarrow 3$ symmetric:
\begin{equation}
M_e M_e^{\dagger} =
\left(
\begin{array}{ccc}
F & A & A \\
A & G & B \\
A & B & G 
\end{array} \right) \ ,
%\eqno(4.13)
\end{equation}
so that $U_L^e$ has a form of a nearly bimaximal mixing.
For simplicity, we assume that $U_L^e$ is given by the
full bimaximal mixing form
\begin{equation}
U^e_L= (U^e_L)^{T} =
\left(
\begin{array}{ccc}
0 & -\frac{1}{\sqrt2} & \frac{1}{\sqrt2} \\
-\frac{1}{\sqrt2} & \frac{1}{2} & \frac{1}{2} \\
\frac{1}{\sqrt2} & \frac{1}{2} & \frac{1}{2} 
\end{array} \right) \ ,
%\eqno(4.14)
\end{equation}
which demands the constraint $F=B+G$ on the matrix (4.13).
Then, the eigenvalues $D^2_e = {\rm diag} 
(m^2_e , \ m^2_{\mu} , \ m^2_{\tau})$
are given by
\begin{eqnarray}
m^2_e &=& G-B \ , \nonumber \\
m^2_{\mu} &=& G+B - \sqrt{2}A  \  , \\
m^2_{\tau} &=& G+B + \sqrt{2}A  \  , \nonumber
%\eqno(4.15)
\end{eqnarray}
On the other hand, we assume that $K$ in the initial basis is 
given by the form
(4.8) with $K_{23} = K_{32}$, so that we obtain $a = a^{\prime}$ and
\begin{equation}
\widehat{M}_{\nu} = 2m_0^{-1} 
\left(
\begin{array}{ccc}
-(g^2 - b^2) & a(g-b) & a(g-b) \\
a(g-b) & a^2 - fg & -(a^2 - fb) \\
a(g-b) & -(a^2 - fb) & a^2 - fg 
\end{array} \right) \ .
%\eqno(4.16)
\end{equation}
Note that the mass matrix (4.16) does not include the contributions 
($\xi$-terms) from nonvanishing
sneutrino VEVs because of $K^T = K$. 
The mass matrix (4.16) gives the following 
eigenvalues and mixings:
\begin{eqnarray}
m_{\nu 1} &=& (g-b) \left[\sqrt{9 (g+b)^2 + 2f(g+b) + f^2} - (g+b+f)
\right] m^{-1}_0 \ , \nonumber \\
-m_{\nu 2} &=& -(g-b) \left[\sqrt{9 (g+b)^2 + 2f(g+b) + f^2} +g+b+f
\right] m^{-1}_0 \  , \\
m_{\nu 3} &=& -2 \left[2a^2 - (g+b)f \right] m^{-1}_0 \  , \nonumber
%\eqno(4.17)
\end{eqnarray}
\begin{equation}
\widehat{U}_{\nu} = 
\left(
\begin{array}{ccc}
c_{\nu} & s_{\nu} & 0 \\
\frac{1}{\sqrt2} s_{\nu} & -\frac{1}{\sqrt2} c_{\nu} & -\frac{1}{\sqrt2} \\
\frac{1}{\sqrt2} s_{\nu} & -\frac{1}{\sqrt2} c_{\nu} & \frac{1}{\sqrt2} 
\end{array} \right) \ ,
%\eqno(4.18)
\end{equation}
\begin{equation}
s_{\nu} = \sqrt{\frac{m_{\nu 1}}{m_{\nu 1} + m_{\nu 2}}} \ , \ \ \ 
c_{\nu} = \sqrt{\frac{m_{\nu 2}}{m_{\nu 1} + m_{\nu 2}}} \ ,
%\eqno(4.19)
\end{equation}
so that we obtain
\begin{equation}
\tan^2 \theta_{solar} = \frac{m_{\nu 1}}{m_{\nu 2}}  \ ,
%\eqno(4.20)
\end{equation}
together with $\sin^2 2 \theta_{atm} = 1$ and
$|U_{13}|^2 = 0$.
For a further simple case with $f=0$, which demands
\begin{equation}
K_{23} = K_{32} = \frac{1}{2}(K_{33} + K_{22})  \ ,
%\eqno(4.21)
\end{equation}
we obtain $m_{\nu 1} = m_{\nu 2}/2 =2(g^2 -b^2) m_0^{-1}$,
so that
\begin{equation}
\tan^2 \theta_{solar} = \frac{1}{2} \ ,
%\eqno(4.22)
\end{equation}
\begin{equation}
R \equiv \frac{\Delta m^2_{21}}{\Delta m^2_{32}} = \frac{3}{4}
\frac{(g^2 - b^2)^2}{a^4 - (g^2 - b^2)^2} \ ,
%\eqno(4.23)
\end{equation}
where we have considered
\begin{equation}
a^2 = \frac{1}{8} (K_{33} - K_{22})^2 \gg g^2 - b^2 = K_{11}^2
(K_{33} + K_{22})^2 \ .
%\eqno(4.24)
\end{equation}
The result (4.22) is favorable to the recent solar \cite{solar}
and KamLAND data \cite{kamland}.
By using the best fit values $\Delta m^2_{solar} =7.2 \times
10^{-5}$ eV$^2$ \cite{solar,kamland} and 
$\Delta m^2_{atm} =2.4 \times 10^{-3}$ eV$^2$ \cite{atm,K2K},
we obtain
\begin{equation}
\frac{m_{\nu 2}}{m_{\nu 3}} = \frac{|g^2-b^2|}{a^2} =
\sqrt{ \frac{4R}{3+4R}} =0.20 \ ,
%\eqno(3.25)
\end{equation}
where $R= \Delta m^2_{solar}/\Delta m^2_{atm}$, and
\begin{equation}
m_{\nu 1} = 0.0049 \ {\rm eV}, \ \ 
m_{\nu 2} = 0.0098 \ {\rm eV}, \ \  
m_{\nu 3} = 0.050 \ {\rm eV}, 
%\eqno(4.26)
\end{equation}
where we have used the relation $m_{\nu 1}/m_{\nu 2}=1/2$
and $\Delta m^2_{atm}=3 m^2_{\nu 2}/4$.
Of course, this is only an example, and the result (4.22) is not 
a prediction
which is inevitably driven from the general form of $M_{\nu}$.

%%%%%%%%%%%%%%%%%%%%%%%%%%%%%%%%%%%%%%%%%%%%%%%%%
%\vspace{5mm}
\section{Summary} 
\label{sec:5}

  In conclusion, within the framework of a SUSY GUT model, 
we have proposed an 
$R$-parity violation mechanism which is harmless 
for proton decay and 
investigated a general form of the neutrino mass matrix $M_{\nu}$. 
As we have
given in Eq.~(3.12), the form of $M_{\nu}$ is described 
in terms of the matrix
$K$ defined in Eq.~(3.4). 
(i) If $K^T = K$, the VEVs of sneutrinos 
are exactly zero, $\langle \widetilde{\nu}_i \rangle =0$, 
in the arbitrary basis, so that 
$M_{\nu}$ is given only by the radiative contributions. 
(ii) If $\widehat{K}$ 
is $2 \leftrightarrow 3$ symmetric, then $\widehat{M}_{\nu}$ is also 
$2 \leftrightarrow 3$ symmetric, so that $\widehat{M}_{\nu}$ can predict 
$\sin^2 2\theta_{atm} =1$ and $|U_{13}|^2 =0$.

In order to demonstrate that the general form indeed has a 
phenomenologically favorable parameter range, we have given 
a simple example
of $K$ and $M_e M_e^{\dagger}$ in the last part of the section 4. 
Although such
a simple form of $K$, (4.8), with the constraint (4.23) is likely, 
the 
investigation of the origin of the possible form $K$ will be 
our next task. 
The 
purpose of the present paper is not to give a special model for neutrino
phenomenology, and it is to demonstrate that it is indeed possible 
to build a
neutrino mass matrix model with $R$-parity violation, i.e. 
without a seesaw
mechanism, even if the model is within a framework of GUT.

The present model has assigned Z$_3$ quantum numbers to the
superfields differently from those in the previous model 
\cite{SUSY-Koide} with  
$\overline{5}_L \leftrightarrow \overline{5}^{\prime}_L$ mixing: 
we have been able to assign the same Z$_3$ quantum number to 
the matter fields $\overline{5}_L$ and $10_L$ 
(and also to $\overline{5}^{\prime}_L$ and $5^{\prime}_L$). 
This re-assignment will give fruitful potentiality for a further 
extension of the present model.

%%%%%%%%%%%%%%%%%%%%%%%%%%%%%%%%%%%%%%%%%%%%%%%%
\vspace{4mm}

\centerline{\large\bf Acknowledgments}
The author would like to thank J.~Sato for helpful
conversations.
This work was supported by the Grant-in-Aid for
Scientific Research, the Ministry of Education,
Science and Culture, Japan (Grant Number 15540283).

%%%%%%%%%%%%%%%%%%%%%%%%%%%%%%%%%%%%%%%%%%%%

%\end{multicols}

\end{document}